\documentclass[prl,showpacs,twocolumn]{revtex4}
\usepackage{graphicx}
\usepackage{epstopdf}
\usepackage{amsmath}
\usepackage{multirow}
\usepackage{rotating}
\usepackage{xcolor}
\usepackage{natbib}
\usepackage{array}
\begin{document}

\title{Comparing XMCD and DFT with STM spin excitation spectroscopy for Fe and Co adatoms on Cu$_{2}$N/Cu(100)$\footnote{Physical Review B 92, 184406 (2015);
 DOI: http://dx.doi.org/10.1103/PhysRevB.92.184406}$}

\author{M. Etzkorn$^{1}\footnote{m.etzkorn@fkf.mpg.de}$}
\author{C.F. Hirjibehedin$^{2}$}
\author{A. Lehnert$^{1}$}
\author{S. Ouazi$^{1}$}
\author{S. Rusponi$^{1}$}
\author{S. Stepanow$^{3}$}
\author{P. Gambardella$^{3}$}
\author{C. Tieg$^{4}$}
\author{P. Thakur$^{4}$}
\author{A.I. Lichtenstein$^{5}$}
\author{A.B. Shick$^{6}$}
\author{S. Loth$^{7}$}
\author{A.J. Heinrich$^{8}$}
\author{H. Brune$^{1}$}

\affiliation{$^{1}$Institut de Physique de la Mati\`{e}re Condens\'{e}e, Ecole Polytechnique F\'{e}d\'{e}rale de Lausanne, CH-1015 Lausanne, Switzerland}
\affiliation{$^{2}$London Centre for Nanotechnology, Department of Physics \& Astronomy, Department of Chemistry, UCL, London WC1H 0AH, United Kingdom}
\affiliation{$^{3}$Department of Materials, ETH Z\"{u}rich, H\"{o}nggerbergring 64, CH-8093 Z\"{u}rich, Switzerland}
\affiliation{$^{4}$European Synchrotron Radiation Facility, B.P. 220, 38043 Grenoble, Cedex, France }
\affiliation{$^{5}$University of Hamburg, Jungiusstrasse 9, 20355 Hamburg, Germany}
\affiliation{$^{6}$Institute of Physics, ASCR, Na Slovance 2, CZ-18221 Prague, Czech Rep.}
\affiliation{$^{7}$Max Planck Research Group - Dynamics of Nanoelectronic Systems, Center for Free-Electron Laser Science, Luruper Chaussee 149, 22761 Hamburg, Germany}
\affiliation{$^{8}$IBM Research Division, Almaden Research Center, San Jose, CA95120, USA}

\date{\today}

\begin{abstract}
We report on the magnetic properties of Fe and Co adatoms on a Cu$_{2}$N/Cu(100)-$c(2 \times 2)$ surface investigated by x-ray magnetic dichroism measurements and density functional theory (DFT) calculations including the local coulomb interaction. We compare these results with properties formerly deduced from STM spin excitation spectroscopy (SES) performed on the individual adatoms. In particular we focus on the values of the local magnetic moments determined by XMCD compared to the expectation values derived from the description of the SES data.The angular dependence of the projected magnetic moments along the magnetic field, as measured by XMCD, can be understood on the basis of the SES Hamiltonian. In agreement with DFT, the XMCD measurements show large orbital contributions to the total magnetic moment for both magnetic adatoms. 
\end{abstract}
\maketitle

Single magnetic atoms adsorbed on nonmagnetic surfaces enable fundamental insights into the origins of magnetic properties~\cite{bru09, rau14} and have the potential of long relaxation times of magnetic states that can be used for quantum information processing or storage~\cite{loth12, miy13}. To investigate such systems, x-ray magnetic circular dichroism (XMCD) is a well established technique that allows us to determine element specifically the spin ($m_S$), dipole ($m_D$), and orbital ($m_L$) moments and their anisotropies~\cite{tho92,car93}. The dipole moment reflects the inhomogeneous spin density distribution within the atom which influences the transition matrix elements in the XMCD sum rules~\cite{car93}. Compared to atomic dimensions XMCD is a spatially averaging technique. However, due to its high sensitivity, ensembles of individual adatoms can be probed at coverages where the adatoms are sufficiently distant from one another such that their mutual interactions become negligible~\cite{gam02, gam03}.

For individual atoms or molecules, spin-excitation spectroscopy (SES) with a scanning tunneling microscope (STM) enables access to magnetic properties like the gyromagnetic ratio ($g$) and magnetic anisotropies~\cite{hei04, hir06, hir07, ott08, che08, tsu09, lot10np, lot10s, kha10, kha11, kah12, Wulf12, hei13, obe14, don13, kha13, Bryant13, rau14, Choi14, Spinelli14, Bergmann15, Yan15}. These studies have been made on magnetic impurities with very different degrees of hybridization with the substrate, ranging from direct contact with metals, semiconductors, and superconductors, to adatoms adsorbed on thin insulating films or graphene decoupling layers. In the vast majority of cases, the SES results were discussed on the basis of an effective spin Hamiltonian using an atomic picture that has the form~\cite{hir07}:
\begin{equation}
  \hat{H} = g \mu_{B} \hat{\vec{S}} \cdot \vec{B} + D \hat{S}_{z}^{2} + E (\hat{S}_{x}^{2} - \hat{S}_{y}^{2}).
\end{equation}
This description is well established in the field of electron spin resonance (ESR)~\cite{abr70, Rudowicz15} and molecular magnets~\cite{wer10} and typically used for systems where the local moments are well protected from hybridization with the conduction electrons of a metal. STM studies are naturally limited to conductive systems in which the electronic transport through the adatoms or molecules to the substrate is significant. This may alter the properties of the system and questions the use of an atomic description for its excitations motivating alternative models~\cite{kha11}. Nevertheless, the described spin Hamiltonian reproduces the energies and amplitudes of the observed SES conductance steps in detail~\cite{hir06, Ternes15}.

It seems reasonable to assume that on the energy scale of the spin excitations the moments of the impurity and its surrounding are sufficiently strongly coupled to behave as one single magnetic moment. In this case SES detects the integrated spin moment with a weighted average anisotropy. Contrary XMCD, measuring intra-atomic transitions, only detects the moments carried by $d$-electrons localized on the impurity itself, due to the nature of the core excitation. In an external magnetic field it is however the total magnetic moment that determines the alignment of the moment of the impurity along the field direction. Therefore SES and XMCD are expected to probe different quantities that nevertheless show the same dependence on strength and orientation of the magnetic field and the magnetic anisotropy of the system.

Here we investigate Fe and Co monomers on Cu$_{2}$N/Cu(100)-c$(2 \times 2)$ with XMCD and spin-resolved density functional theory (DFT) and compare it to published STM-SES data~\cite{hir07,ott08}. On one hand we find that the angular dependence of the XMCD is in good agreement with the extrapolated angular dependence from the SES results. On the other hand we find significant orbital magnetic moments and comparably small spin moments localized on the adatom. This illustrates the influence of the adatoms´ orbital moment and the induced moments of the environment on the SES spectra.

\subsection{Experimental details:}

\begin{figure}
\begin{center}
\includegraphics[width = 6.0 cm]{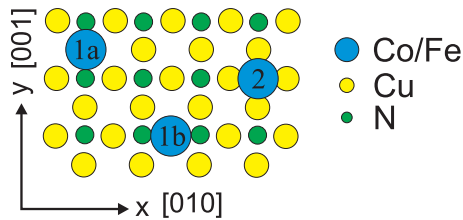}
\end{center}
\caption{Topview of the three adsorption sites populated by Fe and Co adatoms on Cu$_{2}$N/Cu(100)-$c(2 \times 2)$. 1a and 1b denote atop Cu sites with two orthogonal in-plane orientations of N neighbors. 2 denotes the atop N site having fourfold in-plane symmetry.}
\label{sites}
\end{figure}

To properly interpret the XMCD ensemble measurements it is important to know the adsorption sites of the adatoms on Cu$_2$N and their occupation statistics, which we obtained from additional STM experiments. For both Fe and Co, we find the three adsorption sites depicted in Fig.~\ref{sites}. The two chemically identical sites (1a, 1b) are atop Cu atoms and distinguish themselves by the orientation of the two next nearest N neighbors. The third site (2) is atop N atoms and thus with fourfold in-plane symmetry. The fourth possible site, the fourfold Cu hollow, is not occupied.

The two chemically inequivalent sites are distinctly different concerning their spectroscopic features in STS. The atop Cu sites show the spin excitations discussed in literature~\cite{hir07, ott08}, while adatoms atop N display no low energy spectroscopic features. We find that $66 \pm 3$~\% of the Fe atoms deposited on the Cu$_2$N show the SES spectra typical for the atop Cu sites, in excellent agreement with statistical adsorption. $23 \pm 2$~\% of the atoms show no spectroscopic features as typical for atop N sites. For about 10~\% of the adatoms found on the surface we could not identify the adsorption site from the STS spectra. For simplicity, we consider an equal share of all three possible adsorption sites as expected for statistical adsorption. Based on our experience with Co on Cu$_2$N, we know that the same adsorption sites as for Fe are populated with similar occurrence. We therefore also assume that Co adatoms exhibit equal adsorption site partition.
A recent study has shown that for Co atop Cu the anisotropy changes for adatoms located in the center of exceptionally large Cu$_2$N islands compared to adsorption sites on either the rim region of such islands or on islands with typical dimensions below $10~nm$~\cite{obe14}. We neglect such deviations due to the very small relative abundance of those adatoms. Co adsorbed atop Cu sites is a Kondo system~\cite{ott08}. However its Kondo temperature of $T_{\rm K} = 2.6$~K is lower than the XMCD measurement temperature of 8~K, and therefore the Kondo screening of this species is expected to be very weak and will be equally neglected in the following.

The XMCD spectra were obtained in the total electron yield mode at the beam line ID08 of the European synchrotron radiation facility (ESRF). The x-ray beam and magnetic field are parallel and form an angle $\theta$ with the sample normal [see inset Fig.~\ref{Co}(a)]. All x-ray measurements have been performed with the sample at the base temperature of this set up of $8$~K. The sample has been oriented such that the projection of the in-plane magnetic field direction is almost parallel (within 10$^{\circ}$) to the crystalline [001] direction.
For the sum rule analysis ~\cite{tho92,car93} one typically normalizes the measured dichroism to the total isotropic absorbtion to determine the absolute values of the magnetic moments. We have used the common approximation $I^{iso} =\frac{3}{2}(I^{\mu^{+}}+I^{\mu^{-}})$, where $I^{\mu^{\pm}}$ denote the intensity of the two helicities of the incoming circularly polarized x-rays. In linear dichroism measurements we find that the energy integrated dichroism of Fe is almost vanishing while for Co it is not (see also the appendix). From these findings we conclude that the above approximation is valid for Fe, while for Co we will under(over)estimate the values along $\theta = 0^{\circ} (70^{\circ})$. The resulting error is included in the given error bars. The Cu(100) substrate was prepared by cycles of Ar sputtering at 300~K followed by 10~min annealing at 820~K. Cu$_{2}$N/Cu(100)-c$(2 \times 2)$ was grown by N sputtering (1~keV, $p_{N_2} = 2 \times 10^{-5}$~mbar, 4~min, 300~K) and annealing to about 600~K for 2~min. These conditions yield a self-limited coverage very close to a complete Cu$_2$N monolayer (ML)~\cite{ell01}. The quality of the Cu$_{2}$N films have been checked with STM and low energy electron diffraction. Fe and Co have been evaporated from high purity rods onto the sample held at below 10~K, a temperature where thermal diffusion is inhibited resulting in statistical growth~\cite{bru98}. The deposition flux has been calibrated from x-ray absorption intensities and cross-checked by room temperature STM measurements of Fe and Co deposited on Cu(100). The coverage in this paper is given in monolayers (ML) of adsorbates with respect to the underlying Cu(100) surface, \textit{i.e.} 1~ML corresponds to one Fe or Co atom per Cu-surface atom.

\subsection{Density functional theory:}

\begin{figure}
\begin{center}
\includegraphics[width = 6.0 cm]{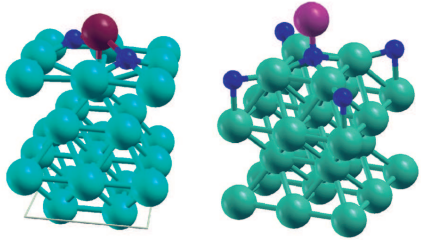}
\end{center}
\caption{Supercell used in the DFT calculations for on-Cu adsorption sites (type 1a/b) and for on-N sites (type 2). Cu atoms are displayed in light blue, N in dark blue, adatoms in red.}
\label{ball}
\end{figure}

The DFT calculation of the magnetic moments and of the magnetic anisotropies were performed on a supercell of three Cu(100) layers, a Cu$_2$N-$c (2 \times 2)$ monolayer, and the Fe and Co adatoms followed by four empty Cu layers modeling the vacuum~\protect\cite{shi09}. Figure~\ref{ball} shows ball models of the supercells employed for the adsorbates atop of Cu (a) and atop of N (b). The structure relaxation was performed employing the VASP code~\protect\cite{kre93,kre96,kre99}. The adatom-substrate distance as well as the atomic positions within the Cu$_2$N layer were allowed to relax.

\begin{table}
\begin{ruledtabular}
\begin{tabular}{ccccc}
			& \multicolumn{2}{c}{Co} 		& \multicolumn{2}{c}{Fe} 		\\
			& Cu site 			& N site 	& Cu site 			& N site	\\
\colrule
$D_{\rm calc}$	& 2.69 			& -4.64	& -1.61 			& -1.41	\\
$D_{\rm SES}$	& $2.75 \pm 0.05$	& --		& $-1.55 \pm 0.01$	& -- 		\\
\colrule
$E_{\rm calc}$	& -0.31			& 0		& 0.17			& 0 		\\
$E_{\rm SES}$	& $\approx 0^{*}$	& --		& $0.31 \pm 0.01$	& -- 		\\
\end{tabular}
\end{ruledtabular}
\caption{DFT and SES~\protect\cite{hir07,ott08} values for the axial ($D$) and transverse ($E$) magnetic anisotropy energies in meV. For the atop Cu sites the axes were chosen as in the cited publications. For the atop N sites both Fe and Co have their easy magnetization axis out of plane. The SES results for Co on atop Cu sites would also agree with small nonzero values of E.}
\label{anis}
\end{table}

\begin{table}
\begin{ruledtabular}
\begin{tabular}{cccccc}
			&	& \multicolumn{2}{c}{Co}	& \multicolumn{2}{c}{Fe}	\\
			& 			& Cu site	& N site	& Cu site	& N site	\\
\colrule
			& $m_S$		&  1.75	&  1.63	& 2.78	& 2.70	\\
$m \parallel z$	& $m_L$		&  0.46	&  2.14	& 0.12	& 0.55	\\
			& $7 \, m_D$	& -0.27	& -0.17	& 0.61	& 2.07	\\
\colrule
			& $m_S$		&  1.77	&  1.56	& 2.78 	& 2.68	\\
$m \parallel x$	& $m_L$		&  0.19	&  1.61	& 0.04 	& 0.20	\\
			& $7 \, m_D$	& -1.77	& -1.22	& -1.72	& -0.96	\\
\colrule
			& $m_S$		&  1.77	&  1.56	& 2.76	&  2.68	\\
$m \parallel y$	& $m_L$		&  0.53	&  1.61	& 0.47	&  0.20	\\
			& $7 \, m_D$	&  1.79	& -1.22	& 1.51	& -0.96	\\
\end{tabular}
\end{ruledtabular}
\caption{Calculated spin $m_S$, orbital $m_L$, and dipole $7 \, m_D$ magnetic moments of $d$ states for Co and Fe on the two adsorption sites for the three spin quantization axes. For the atop Cu site the axes correspond to type 1a in Fig~\ref{sites}, \textit{i.e.}, $x$ and $y$ refer to the vacancy and to the N direction, respectively. Note that for reasons of consistency within this paper the axes are different compared to Refs.~\protect\cite{hir07,ott08}. The units are Bohr magnetons. $U = 2(2)$~eV and $J$ = 0.91(0.84) eV for Co (Fe) have been used in LSDA+$U$ calculations.}
\label{m}
\end{table}

The magnetic properties of the 3$d$ adsorbates are strongly influenced by relativistic and electron correlation effects. The former are accounted for by implementing the local spin-density approximation (LSDA) in the relativistic version of the full-potential linearized augmented plane-wave method (FP-LAPW)~\cite{wim81} in which spin-orbit coupling (SOC) is included in a self-consistent second-variational procedure~\cite{shi97}. To account for the electron correlation we used LSDA+$U$ that adds an energy of a multiorbital Hubbard type and subtracts a double-counting term for the part of the electron-electron interaction energy already included in LSDA. In total this gives the relativistic version of the LSDA+$U$ method implemented in the FP-LAPW basis~\cite{shi01} with which we calculated the spin $m_S$, orbital $m_L$, and $7m_D$ dipole magnetic moments for the Fe and Co adatoms at $T$=0. We used first order perturbation theory to derive from the orbital anisotropies theoretical values for the magnetocrystalline anisotropies $K = -\xi/4 \; \Delta m_L$, where $\xi$ denotes the SOC constant of the $d$ states~\cite{bru89}. We took $\xi_{\rm Co} = 79$~meV and $\xi_{\rm Fe} = 65$~meV. In order to compare these calculated anisotropy values with the numbers derived from the SES measurements, we renormalized $K$ by $S^2$ with $S = 3/2$ for Co~\cite{ott08} and $S = 2$ for Fe~\cite{hir07} to compute the $D$ and $E$ values shown in Table ~\ref{anis}. We have made these calculations with different flavors of LSDA+$U$ (the fully localized and around-mean-field limits) and different values of $U$. We have then chosen the one that had best agreement with the anisotropy values measured with SES on the atop Cu binding sites. The comparison with the XMCD data is therefore performed without additional fitting parameters in the theory.
For atop N sites no spin excitations are observed, thus no information on the magnetic properties of the adatom on this adsorption site can be obtained from SES. Therefore we rely entirely on theory for these values. Fortunately, the most important quantity for the simulations, the calculated anisotropy, was always found to be strongly out-of-plane independent of the flavor of the LSDA+U. The DFT results for the magnetic moments along out-of-plane ($z$) and in-plane directions are shown in Table~\ref{m}.

The calculated spin moments on both magnetic adatoms are much smaller than the values of the effective SES spin operator~\cite{hir07,ott08}. In addition, large orbtial magnetic moments are found. Note that the calculated spin moments are rather isotropic compared to the orbital moments. Finally, we find that $\sum m_D \neq 0$, which is in agreement with previous studies of low dimensional systems with large spin-orbit coupling terms~\cite{ede03b, ede03}.

\subsection{XMCD measurements}

\begin{figure}
\begin{center}
\includegraphics[width = 6.0 cm]{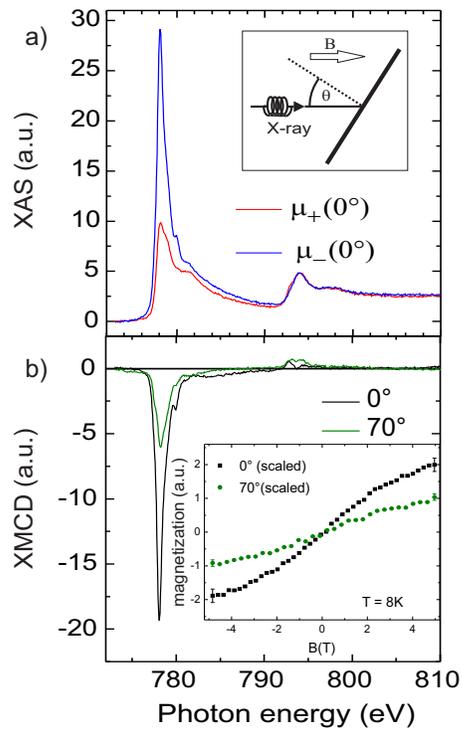}
\end{center}
\caption{a) X-ray absorption spectra (XAS) for 0.02~ML Co on Cu$_{2}$N/Cu(100)-c$(2 \times 2)$ for the two photon helicities $\mu_{+}$ and $\mu_{-}$ with respect to \textit{\textbf{B}}, ($\theta = 0^{\circ}$, $B = 5$~T, $T = 8$~K). The substrate background has been subtracted. The inset shows the measurement geometry. b) XMCD normalized to the total absorption intensity for the respective angles. The inset shows magnetization curves measured by the XMCD $L_3$ peak heights. The magnetization curves have been scaled to the measured $(m_S + 7 \, m_D)_{0^{\circ}} / (m_S + 7 \, m_D)_{70^{\circ}}$ ratio at $B = 5~T$.}
\label{Co}
\end{figure}

\begin{figure}
\begin{center}
\includegraphics[width = 5.8 cm]{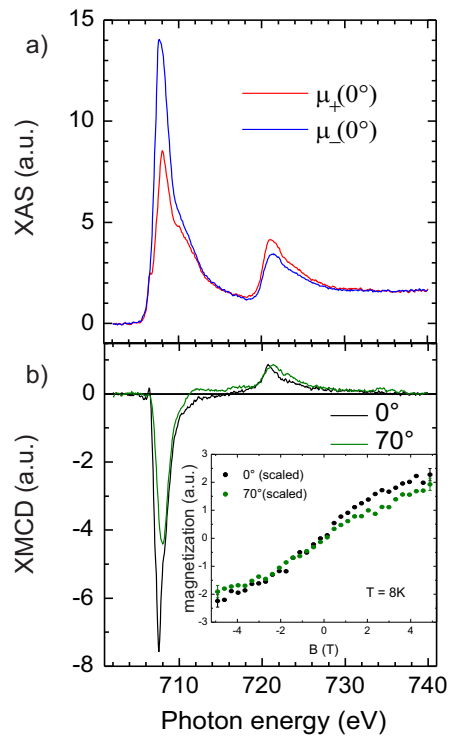}
\end{center}
\caption{a) Background corrected XAS for 0.02~ML Fe on Cu$_{2}$N/Cu(100)-c$(2 \times 2)$ ($\theta = 0^{\circ}$, $B = 5$~T, $T = 8$~K). b) Corresponding angular dependent XMCD. The inset displays magnetization curves obtained as in Fig. 3(b).}
\label{Fe}
\end{figure}

The x-ray absorption spectra (XAS) of 0.02 ML Co and Fe on Cu$_{2}$N/Cu(100)-c$(2 \times 2)$ are shown in Figs.~\ref{Co}(a) and \ref{Fe}(a). The angle-dependent XMCD spectra and the XMCD magnetization curves are displayed in Figs.~\ref{Co}(b) and \ref{Fe}(b). The XAS of both elements show rather narrow spectral features indicating that the electronic coupling of the adatoms to the metallic states of the Cu(100) substrate is significantly reduced by the Cu$_{2}$N layer. This is in agreement with the reported low conductance through the nitride monolayer decreasing the spin scattering between the magnetic atom and substrate electrons~\cite{lot10np}.

As visible from the magnetization curves in the insets of Fig.~\ref{Co}(b) and \ref{Fe}(b), both systems cannot be saturated with the available maximum field of 5~T at the sample temperature of 8~K. Therefore the absolute saturation values of spin and orbital moments cannot be determined from XMCD alone. However, we can determine the anisotropy of the spin and dipole moment from the sum rules~\cite{tho92, car93} as $(m_S + 7 \, m_D)_{0^{\circ}} / (m_S + 7 \, m_D)_{70^{\circ}}$ and the relative contribution of the orbital moment to the sum of spin and dipole moment $m_L / (m_S + 7 \, m_D)$. These are summarized in Table~\ref{Sumrule}. For both Fe and Co the orbital contributions to the magnetic moments are rather large and exhibit a significant anisotropy.

\begin{table}
\begin{ruledtabular}
\begin{tabular}{ccc}
						& Co 			& Fe	\\
\colrule
$(m_S + 7 \, m_D)_{0^{\circ}} / (m_S + 7 \, m_D)_{70^{\circ}}$	& $1.82 \pm 0.37$			& $1.22 \pm 0.18$	 \\
\colrule
$m_L / (m_S + 7 \, m_D)(0^{\circ})$	& $0.62 \pm 0.10$	& $0.33 \pm 0.05$	\\
\colrule
$m_L / (m_S + 7 \, m_D)(70^{\circ})$ & $0.41 \pm 0.11$	& $0.15 \pm 0.06$ \\
	\\
\end{tabular}
\end{ruledtabular}
\caption{Ratio of the sum of spin and dipole moment for $\theta = 0^{\circ}$ and $ 70^{\circ}$ as well as ratio of orbital moment to the sum of spin and dipole moment from the sum rules analysis of the XMCD data shown above.}
\label{Sumrule}
\end{table}


\subsection{Discussion:}

To compare our results to the SES data, it is important to recall some of the fundamental properties of the spin Hamiltonian. Within an atomic picture the magnetic properties of the adatom can be described by a spin and orbital moment, coupled via spin-orbit coupling, in the presence of a crystal (or ligand) field~\cite{pry50, abr70}. For 3$d$ impurities the latter can be expected to be the leading energy term. To describe the spin excitations of the system, for example in ESR, a perturbative approach yields the effective spin Hamiltonian in which the orbital moments are included in an (anisotropic) $g$ value different from 2, within the anisotropy constants and within the multiplicity of the moment operator~\cite{pry50, abr70}. The effective spin operator enters in two respects. Its value reflects the possible transitions in the excitation spectrum of the magnetic system~\cite{abr70,Rudowicz15}; it also enters in the Zeeman term, and therefore the product of gS reflects the magnetic susceptibility~\cite{abr70,Rudowicz15}.

In this respect it is interesting to note that SES is rather sensitive to the multiplicities of the operator, because the spin excitation spectra change significantly between $\hat{S}$ being integer or not. From our DFT calculations we derive the number of $d$ electrons localized in the adatoms to be very close to 7 (6) for Co (Fe) (within 1\% for both),which seems to suggest that the value of the spin operator is simply given by the spin of the free atom, but of course the presence of spin orbit coupling as well as of the hybridization to the surface strongly alter this picture, as can for example be seen from the values of the magnetic moments given by the DFT calculations.

Concerning the Zeeman energy we have to compare the magnetic moments to the values of $g \hat{S}$ (with a g value different from 2) determined in SES. The SES values are $g \hat{S} = 3.29 \pm 0.13 \; \mu_{B}$ for Co, and $4.22 \pm 0.10 \; \mu_{B}$ for Fe~\cite{hir07, ott08}, which is significantly larger than the spin moments of 1.77 $\mu_{B}$ and 2.78 $\mu_{B}$ calculated by DFT. Considering the calculated total spin moment in the supercell (2.42~$\mu_{B}$ for Co and 3.58~$\mu_{B}$ for Fe) and adding the orbital moments [up to 0.53 (0.47) $\mu_{B}$ for Co (Fe)], one finds values of the total moment that are close to the values suggested from the spin Hamiltonian description, though in the calculations even these moments remain below the SES values.

We now combine the DFT results with the ones from SES to establish the closest possible comparison with XMCD. As mentioned before, in the XMCD measurements the moments are not saturated. Thus the different species have different degrees of saturation and therefore contribute differently to the ensemble average values measured in XMCD. The spin Hamiltonian described above can be used to calculate the degree of saturation under the experimental conditions used in the XMCD measurements. For given values of $\hat{S}$, $g$, $D$, and $E$ and magnetic field, we calculate the eigenstates $\hat{S}_i$ and their energies $\xi_i$ from the Hamiltonian. The expectation value of the spin operator for a given temperature is then given by a Boltzmann weighted average of these eigenstates: $\langle\hat{S}\rangle = \frac{\sum_{i = 1}^{2 S + 1} \hat{S}_i e^{- \xi_i /     k_{\rm B} T}} {\sum_{i = 1}^{2 S + 1} e^{-\xi_i / k_{\rm B} T}}.$ \noindent The resulting degree of saturation $\langle\hat{S}\rangle/S$ is a measure of the contribution of the moments of this particular species to the ensemble average. We thus weight the $m_{L}$ and $m_{S} + 7 \, m_D$ derived from DFT with the $\langle\hat{S}\rangle/S$ value for each species. Finally we average over the three adsorption sites to derive the average value of $m_{L}$ and $m_{S} + 7 \, m_D$ that can be compared to the XMCD results. One ambiguity using the DFT values in the spin Hamiltonian is the question of which moment operator to use, in particular for atop N sites, where no SES data is available. To start with, we have used the spin operator found in the SES measurements for the atop Cu sites also for the atop N site. Using the values of $g$, $D$, and $E$ from SES for the atop Cu sites and from DFT for the atop N site, as well as the spin and dipole moment of both adsorption sites from DFT, we calculate from the spin Hamiltonian $(m_S + 7 \, m_D)_{0^{\circ}} / (m_S + 7 \, m_D)_{70^{\circ}} = 1.68 (1.36)$ for Co (Fe) at $B = 5~T$ and $T = 8~K$. For the relative size of the orbital moment this model predicts for Co $m_L / (m_S + 7 \, m_D)$ = 0.79 (0.37)for $0^{\circ}(70^{\circ})$ and for Fe 0.10 (0.08). Thus, five out of the six values from this analysis are matching the results extracted from the XMCD measurements or are very close. Only the ratio of orbital to spin moment for Fe at $\theta = 0^{\circ}$ shows significant deviations between the XMCD measurements and the analysis based on the spin Hamiltonian.  We consider this level of agreement to be very good taking into account the uncertainties of some of the values used. We note that this model is not a unique solution in the sense that other combinations of parameters (anisotropies and moments) will reproduce the XMCD results at the same or an even better level. Nevertheless finding such a level of agreement between XMCD, DFT, and SES using one common description is rather satisfactory. As discussed in the appendix this agreement is also found for different values of the initial spin operator used.


As discussed above, the values of the magnetic moments found in the Zeeman term of the SES description can to a good extent be accounted for in the DFT calculations by considering all moment contributions, spin and orbital moments on the adatom, as well as induced moments of the surrounding. To this end, we determine the (unsaturated) moments localized on the adatoms from XMCD using the number of $d$ holes from the calculations. We find for Co $(m_S + 7 \, m_D) = 1.44 \pm 0.14$ $(0.79 \pm 0.14)$ for $\theta = 0^{\circ}(70^{\circ})$. For Fe these values are $1.69 \pm 0.18$ $(1.39 \pm 0.14)$.

If the value of the spin operator would reflect the size of the spin moment localized on the adatom, we would expect values of $(m_S + 7 \, m_D)$ = 1.75(1.02) for Co and 2.82(2.05) for Fe from our estimations using the spin Hamiltonian~\footnote{Here we have again followed the above routine to calculate the degree of saturation for each adsorption site under the experimental conditions. However, instead of multiplying these values with the DFT values of the spin moments, we have now multiplied them with 3(4)~$\mu_{B}$ for Co(Fe). In addition, we have assumed the dipole moment to linearly increase with spin moment used. If we take the DFT values of the dipole moments directly the deviations are even more significant.}. In both cases, in particular for Fe, the calculated values are far larger than the measured ones. The above discrepancy is a clear indication that the orbital moments and the induced moments in the neighboring Cu and N atoms also contribute to the moment operator, as also suggested in DFT. Though it is well known that for example in the ESR the moment operator should be considered as an "effective spin" operator only~\cite{abr70, Rudowicz15}, the above comparison clearly illustrates that this also holds true for the interpretation of SES data.

As pointed out before, neither Fe nor Co adatoms atop N sites show spectroscopic features that can be related to spin excitations in SES. Using the above Hamiltonian, that explains both energies and transition intensities measured in SES spectra for the atop Cu sites, the absence of spin excitations can only be explained by either a small relative cross section of the excitation or with the absence of magnetic moments on this adsorption site. Following the above routine, but assuming no magnetic moment on the atop N site, results in the following values for Co (Fe): $(m_S + 7 \, m_D)_{0^{\circ}} / (m_S + 7 \, m_D)_{70^{\circ}}$ = 1.11 (0.36); $m_L / (m_S + 7 \, m_D)(0^{\circ})$ = 0.31 (0.04); $m_L / (m_S + 7 \, m_D)(70^{\circ})$ = 0.22 (0.07). As can be seen from these numbers, the level of disagreement with the XMCD values is significant, even for instance predicting the wrong sign of anisotropy for Fe. We conclude that both Fe and Co have significant magnetic moments on the atop N adsorption site. In return, this raises the question of what determines the cross section of spin excitations in SES experiments~\cite{lounis14}.

In conclusion, we have presented a combined XMCD and DFT study of the magnetic properties of Fe and Co monomers on Cu$_{2}$N, which we compare to published SES results of these systems. We are able to describe the results of these three very different approaches on one common basis. In particular the anisotropy constants determined with SES are in qualitative agreement with the angular dependence of the spin moments determined from XMCD. Both XMCD and DFT show that both adatoms have rather large orbital magnetic moments. The above mentioned agrement can only be found when considering the contributions of those orbital moments, as well as of the induced moments of the surrounding. In addition our results indicate large magnetic moments on the Co and Fe adatoms atop N sites albeit no magnetic excitations have yet been observed on these species.

\begin{acknowledgements}
C. F. H. acknowledges funding from EPSRC grant EP/D063604/1.  A.B.S. acknowledges support from the Czech Science Foundation (GACR)Grant No. 15-07172S, and access to computing and storage facilities of the National Grid Infrastructure MetaCentrum was provided under the program LM2010005.
\end{acknowledgements}

\section{Appendix:}

Within the appendix we give further details on the DFT calculations, the used spin Hamiltonian and the influence of the value of the moment operator on the $\frac{[(m_{S}+m_{7D})(0^{\circ})]}{[(m_{S}+m_{7D})(70^{\circ})]}$ values to which the XMCD measurements are compared. In addition we show x-ray adsorbtion spectra measured with linear polarized light.



\subsection{DFT calculations}

\begin{figure}[h]
\centerline{\includegraphics[angle=270,width=0.95\columnwidth,clip]{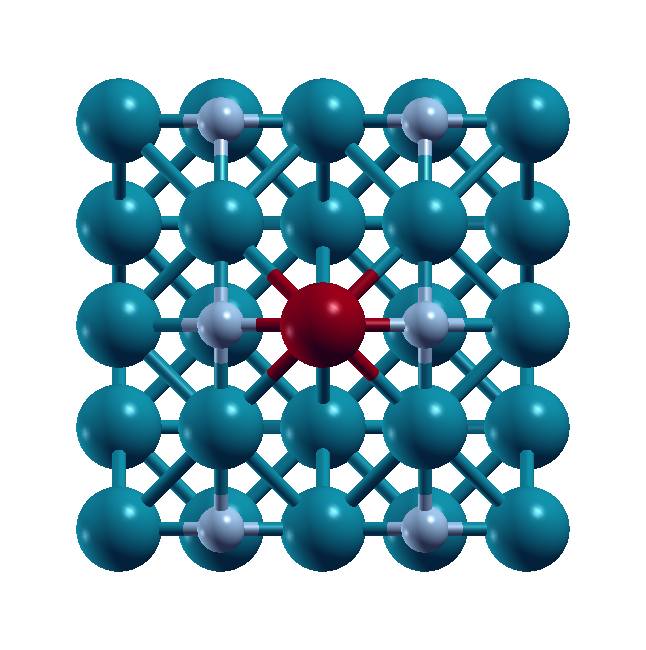}}
\caption{Top view of Model-III with the adsorption site atop Cu with a supercell of Cu$_8$N$_4$] and 3 Cu(100) substrate layers.}
 \label{fig1}
\end{figure}

In order to check the possible influence of the size of the model on the obtained results, we altered the lateral extent of the supercell used in the calculations. Figure~\ref{fig1} illustrates an example of a model for an atop Cu adsorbing site corresponding to a coverage of 0.125 ML. The structure
relaxation is performed employing the standard VASP method without
spin-orbit coupling (SOC) and making use of the generalized gradient
approximation.

Next, we use the relativistic version of the full-potential
linearized augmented plane-wave method (FP-LAPW), in which
spin-orbit (SO) coupling is included in a self-consistent
second-variational procedure. We use
the rotationally-invariant relativistic LSDA+U method with the ``fully
localized limit" (FLL) form for the double-counting correction.
The values for the Coulomb $U$ = 3 eV and exchange
$J$ = 0.9 eV for Co and $U$ = 3 eV and exchange $J$ = 0.84 eV for Fe
were used in LSDA+U calculations. Note that these values are in the ballpark of commonly accepted $U$ and $J$ for transitional 3$d$ metals~\cite{Sasioglu15}.In these calculations, a set of
$k$ points equivalent to 49 $k$ points in the full 2D BZ is used.

The spin $m_S$, orbital $m_L$, and $7m_D$ dipole magnetic moments for
the $d$ states, for the magnetization along the $x,y,$ and $z$ axes
(the $x$ axis is chosen along the in-plane hollow direction, the
$y$ axis is along the in-plane N chain direction, and $z$-axis is
along the out-of-plane direction, perpendicular to the surface), are
given in Table~I for Co and Fe adatoms atop Cu and N. Small spin and orbital moments are also
induced on neighboring Cu sites and quickly decay away from the
magnetic Co and Fe atoms. In these calculations the charge/spin
density self-consistency is performed for all three ($x,y,z$)
directions of the magnetization.

\begin{table}[htbp]
\begin{ruledtabular}
\begin{tabular}{cccccc}
			&			& \multicolumn{2}{c}{Co}	& \multicolumn{2}{c}{Fe}	 \\
			& 			& Cu site	& N site	& Cu site	& N site	\\
\colrule
			& $m_S$		&  1.79	&  1.76	& 2.93	& 2.80	\\
$m \parallel z$	& $m_L$		&  0.69	&  2.30	& 0.15	& 1.17	\\
			& $7 \, m_D$	& -0.94	& -0.36	& -0.06	& 1.84	\\
\colrule
			& $m_S$		&  1.75	&  1.67	& 2.88 	& 2.81	\\
$m \parallel x$	& $m_L$		&  0.43	&  1.92	& -0.06 	& 0.34	\\
			& $7 \, m_D$	& -1.85	& -1.26	& -1.64	& -0.81	\\
\colrule
			& $m_S$		&  1.86	&  1.67	& 2.89	& 2.81	\\
$m \parallel y$	& $m_L$		&  0.60	&  1.92	& 0.45	& 0.34	\\
			& $7 \, m_D$	&  2.34	& -1.26	& 1.67	&-0.81	\\
\end{tabular}
\end{ruledtabular}
\caption{Calculated spin $m_S$, orbital $m_L$, and  $7 \, m_D$ dipole magnetic moments of $d$ states for Co and Fe on the two adsorption sites for three spin quantization axes. For the 1a(b) site the $x$ axis is along the hollow direction and the $y$ axis is along the N chain. The units are Bohr magnetons. $U = 3$~eV and $J$ = 0.91(0.84) eV for Co (Fe) have been used in LSDA+$U$ calculations. }
\end{table}


The results of the calculations in this bigger supercell do not differ qualitatively from those of the smaller supercell. In addition, in the case of the smaller supercell, we have found no significant dependence of the $d$-shell spin- and orbital moments on the number of layers of the Cu substrate. Therefore, we conclude that the results are representative also for the experimental case of diluted adatoms with a coverage of about 2\% of a ML.

\begin{table}[htbp]
\begin{ruledtabular}
\begin{tabular}{ccccc}
			& \multicolumn{2}{c}{Co} 		& \multicolumn{2}{c}{Fe} 		\\
			& Cu site 			& N site 	& Cu site 			& N site	\\
\colrule
$D_{\rm calc}, U= 3$ eV 	& 1.89	& -3.34	& -1.65		& -3.37	\\
\colrule
$E_{\rm calc}, U=3 $ eV	& 0.39	& 0		& 0.43		& 0 		\\
\end{tabular}
\end{ruledtabular}
\caption{Calculated axial ($D$) and transverse ($E$) magnetic anisotropy energies in meV.
For the Co atop Cu: x-N chain,y-perpendicular, and z-hollow axes are chosen;
for the Fe atop Cu: x-hollow, y-perpendicular, and z-N-chain are chosen;
For the atop N sites x-hollow, y-N chain, and z-perpendicular axes are chosen. S=2 and 1.5
are assumed for Fe and Co, when the energy differences are converted into the anisotropy parameters.}
\end{table}

\subsection{The Spin Hamiltonian:}

At the conditions of the XMCD measurements ($T = 8~K, B = 5~T$) the magnetic moments of the ensemble are not saturated. Since the anisotropy along a given crystallographic axis is different for all three species, each will reach a different degree of saturation and therefore will contribute differently to the measured averaged moments. Thus, we first have to estimate the degree of saturation of each species under the experimental conditions. We have employed the effective spin Hamiltonian used to describe the SES data for this purpose~\cite{hir07}:
\begin{equation}
  \hat{H} = g \mu_{B} \hat{\vec{S}} \cdot \vec{B} + D \hat{S}_{z}^{2} + E (\hat{S}_{x}^{2} - \hat{S}_{y}^{2}).
\end{equation}

This Hamiltonian is an approximation used if the orbital ground state of the system is not degenerate~\cite{abr70}. In our approach, we restrict ourselves to thermal excitations starting from the electronic ground state only, \textit{i.e.} assume only one spin moment and its corresponding anisotropies. In principle thermal excitations may also populate electronic states that have different (spin) moment and anisotropies~\cite{abr70,cor12,rau14}. This is potentially a rather drastic simplification at the temperatures of the XMCD measurements ($T =~8$~K).

\begin{table}
\setlength{\extrarowheight}{2pt}
\begin{tabular}{c c|c|c|c|c|c|}
  \cline{3-6}
  &  & \multicolumn{4}{c|}{Moments on Cu site} \\ \cline{3-6}
  & & S=1, L=0 & S=1, L=1 & S=$\frac{3}{2}$, L=0 & S=$\frac{3}{2}$, L=1 \\ \hline

  \multicolumn{1}{|c|}{\multirow{7}{*}{\begin{sideways} Moments on N site \end{sideways} }}& \multicolumn{1}{|c|} {S=$\frac{1}{2}$, L=1} & 2.61 & 1.71 & 1.63 & 1.58 \\
  \multicolumn{1}{|c|}{} & \multicolumn{1}{|c|}{S=$\frac{1}{2}$, L=2} & 2.75 & 1.80 & 1.72 & 1.66 \\ \cline{2-6}
  \multicolumn{1}{|c|}{} & \multicolumn{1}{|c|}{S=1, L=1} & 2.74 & 1.80 & 1.72 & 1.66 \\
  \multicolumn{1}{|c|}{} & \multicolumn{1}{|c|}{S=1, L=2} & 2.71 & 1.81 & 1.73 & 1.67 \\ \cline{2-6}
  \multicolumn{1}{|c|}{} & \multicolumn{1}{|c|}{S=$\frac{3}{2}$, L=0} & 2.71 & 1.80 & 1.71 & 1.65 \\
  \multicolumn{1}{|c|}{} & \multicolumn{1}{|c|}{S=$\frac{3}{2}$, L=1} & 2.70 & 1.81 & 1.73 & 1.67 \\
  \multicolumn{1}{|c|}{} & \multicolumn{1}{|c|}{S=$\frac{3}{2}$, L=2} & 2.60 & 1.79 & 1.72 & 1.66 \\
  \hline
  \end{tabular}
  \caption{$\frac{[(m_{S}+m_{7D})(0^{\circ})]}{[(m_{S}+m_{7D})(70^{\circ})]}$  for Co. We have used the SES anisotropies for the atop Cu site and the DFT value for the atop N site. In both cases the anisotropies were scaled such that $(D*J^2 = const.)$. The spin and dipole moments have been taken from Table I in the paper and the Land\'{e} factor has been used for g. For comparison the XMCD determined value for Co is $\frac{[(m_{S}+m_{7D})(0^{\circ})]}{[(m_{S}+m_{7D})(70^{\circ})]} = 1.82 \pm 0.37$.}
\end{table}

The first question to be answered in this Hamiltonian is which moment operator should be used. Considering that the DFT calculated magnetic moments are non integer values for the spin and orbital moments there is a range of operators that potentially qualify. To illustrate the dependence of the values discussed in the paper on the choice of the moment operator, we have listed in Tables VI-VII the values calculated for different total moments $J(= S + L) $. For the effective $g$ value we used the Land\'{e} factor $g = \frac{2S+L}{S+L}$ \footnote{To retain consistency we have used this g values also for the case discussed in the paper (pure spin operator). In the paper we have however used the g value determined from SES, which leads to the small deviations between the moment ratios stated in the paper and in the appendix.}. The problem is further complicated by the fact that in the spin Hamiltonian the height of the anisotropy barriers are depending on the value of the moment. To compare different values derived from different moment operators we therefore should scale the $D$ and $E$ parameters so that $D*J^{2} = const.$ (same for $E$) to compare the same magnetic anisotropy energies. The spin excitation energies measured in SES also depend on the anisotropy barriers and the moment operators, however, with a different scaling. For instance in the case of Co atop Cu (with $D > 0$) the excitation energies stay the same for any (integer) moment operator used when $D$ = const. For the atop N site (with $D < 0$) the excitation energies remain constant if $D(2J-1)= const.$. For Fe on the atop Cu site (with negative $D$ and nonzero $E$ term) the terms have again to be scaled differently to get the same excitation energies. Due to the different scaling it is impossible to retain consistency with both SES and DFT values when comparing different moment operators.


\begin{table}
\setlength{\extrarowheight}{2pt}
\begin{tabular}{cc|c|c|c|c|c|}
  \cline{3-6}
  &  & \multicolumn{4}{c|}{Moments on Cu site} \\ \cline{3-6}
  & & S=$\frac{3}{2}$, L=0 & S=$\frac{3}{2}$, L=1 & S=2, L=0 & S=2, L=1 \\ \hline

  \multicolumn{1}{|c|}{\multirow{4}{*}{\begin{sideways} on N site\end{sideways}}}& \multicolumn{1}{|c|} {S=$\frac{3}{2}$, L=0} & 1.45 & 1.27 & 1.30 & 1.25 \\
  \multicolumn{1}{|c|}{} & \multicolumn{1}{|c|}{S=$\frac{3}{2}$, L=1} & 1.51 & 1.33 & 1.37 & 1.31 \\ \cline{2-6}
  \multicolumn{1}{|c|}{} & \multicolumn{1}{|c|}{S=2, L=0} & 1.51 & 1.33 & 1.36 & 1.31 \\
  \multicolumn{1}{|c|}{} & \multicolumn{1}{|c|}{S=2, L=1} & 1.52 & 1.34 & 1.38 & 1.32 \\
  \hline
  \end{tabular}
  \caption{$\frac{[(m_{S}+m_{7D})(0^{\circ})]}{[(m_{S}+m_{7D})(70^{\circ})]}$  for Fe. All parameters were chosen as described in Table VI. The XMCD determined value for Fe is $\frac{[(m_{S}+m_{7D})(0^{\circ})]}{[(m_{S}+m_{7D})(70^{\circ})]} = 1.22 \pm 0.18$. }
\end{table}

Summarizing the above results we find that for any total moment operator that would agree with the SES spectra, \textit{i.e.} being integer for Fe and noninteger for Co, the values measured with XMCD are rather insensitive to the value of the moment operator. This is the case for the operators on both adsorption sites. We can therefore not claim any significant preferences of one operator to be best agreeing to our data. We have checked that this conclusion also remains valid for a scaling of the anisotropy parameters to match the spin exciton energies found in SES measurements.

\begin{figure}
\begin{center}
\includegraphics[width = 8.0 cm]{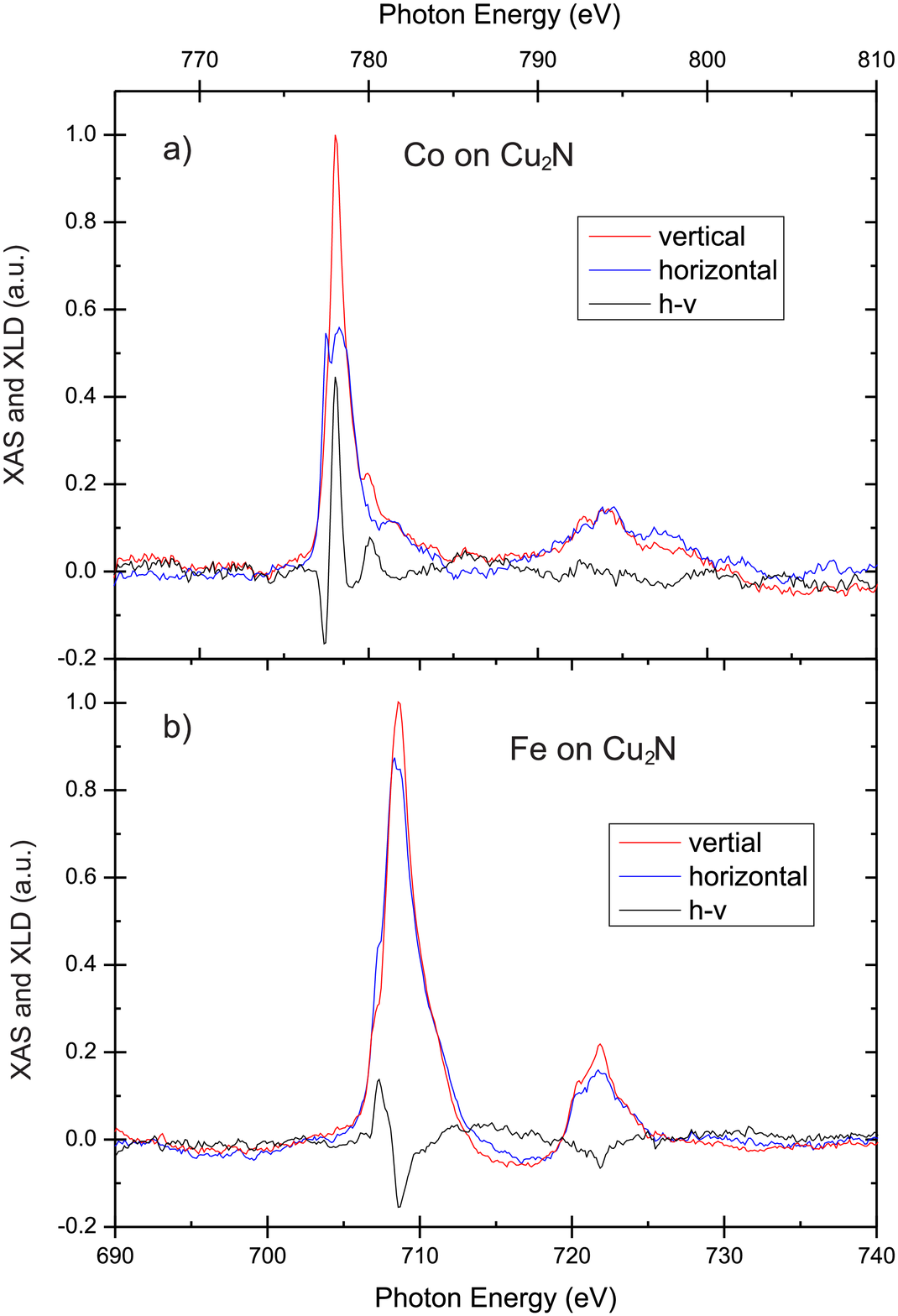}
\end{center}
\caption{XAS spectra recorded with linear polarization and their dichroism for Co (a) and Fe (b) on Cu$_{2}$N both measured at $\theta = 70^{\circ}$ and $B = 0.005~T$. From the shown spectra the background as well as a step function for the non resonant absorption have been subtracted.}
\label{XLD}
\end{figure}

\subsection{X-ray linear dichroism:}

In addition to the circular dichroism experiments, we have measured the x-ray absorbtion for linear light polarization to reveal possible anisotropies in the electronic charge distribution of the $d$ electrons of the impurities (Fig.~6). In principal, together with the XMCD spectra, this information can be used to determine the underlying electronic and spin configuration of the adatoms via multiplet simulations. However, since in our system we average over two chemically different adsorption sites, we have not attempted this task. We would like to show these spectra nevertheless, as they might be very helpful for comparison in further studies.

Integrating the dichroic signal over the full energy range yields values significantly different from zero for Co. This is a clear sign of anisotropic charge distribution between out-of-plane and in-plane orbitals. Such effects are known to result in errors in the approximation $I^{iso} \approx \frac{3}{2}(I^{\mu^{+}}+I^{\mu^{-}})$ typically done in experimental data analysis~\cite{ste10}.

\bibliography{ms_11}

\begin{thebibliography}{50}
\expandafter\ifx\csname natexlab\endcsname\relax\def\natexlab#1{#1}\fi
\expandafter\ifx\csname bibnamefont\endcsname\relax
  \def\bibnamefont#1{#1}\fi
\expandafter\ifx\csname bibfnamefont\endcsname\relax
  \def\bibfnamefont#1{#1}\fi
\expandafter\ifx\csname citenamefont\endcsname\relax
  \def\citenamefont#1{#1}\fi
\expandafter\ifx\csname url\endcsname\relax
  \def\url#1{\texttt{#1}}\fi
\expandafter\ifx\csname urlprefix\endcsname\relax\def\urlprefix{URL }\fi
\providecommand{\bibinfo}[2]{#2}
\providecommand{\eprint}[2][]{\url{#2}}

\bibitem[{\citenamefont{Brune and Gambardella}(2009)}]{bru09}
\bibinfo{author}{\bibfnamefont{H.}~\bibnamefont{Brune}} \bibnamefont{and}
  \bibinfo{author}{\bibfnamefont{P.}~\bibnamefont{Gambardella}},
  \bibinfo{journal}{Surf. Sci.} \textbf{\bibinfo{volume}{603}},
  \bibinfo{pages}{1812} (\bibinfo{year}{2009}).

\bibitem[{\citenamefont{Rau et~al.}(2014)\citenamefont{Rau, Baumann, Rusponi,
  Donati, Stepanow, Gragnaniello, Dreiser, Piamonteze, Nolting, Gangopadhyay
  et~al.}}]{rau14}
\bibinfo{author}{\bibfnamefont{I.~G.} \bibnamefont{Rau}},
  \bibinfo{author}{\bibfnamefont{S.}~\bibnamefont{Baumann}},
  \bibinfo{author}{\bibfnamefont{S.}~\bibnamefont{Rusponi}},
  \bibinfo{author}{\bibfnamefont{F.}~\bibnamefont{Donati}},
  \bibinfo{author}{\bibfnamefont{S.}~\bibnamefont{Stepanow}},
  \bibinfo{author}{\bibfnamefont{L.}~\bibnamefont{Gragnaniello}},
  \bibinfo{author}{\bibfnamefont{J.}~\bibnamefont{Dreiser}},
  \bibinfo{author}{\bibfnamefont{C.}~\bibnamefont{Piamonteze}},
  \bibinfo{author}{\bibfnamefont{F.}~\bibnamefont{Nolting}},
  \bibinfo{author}{\bibfnamefont{S.}~\bibnamefont{Gangopadhyay}},
  \bibnamefont{et~al.}, \bibinfo{journal}{Science}
  \textbf{\bibinfo{volume}{344}}, \bibinfo{pages}{988} (\bibinfo{year}{2014}).

\bibitem[{\citenamefont{Loth et~al.}(2012)\citenamefont{Loth, Baumann, Lutz,
  Eigler, and Heinrich}}]{loth12}
\bibinfo{author}{\bibfnamefont{S.}~\bibnamefont{Loth}},
  \bibinfo{author}{\bibfnamefont{S.}~\bibnamefont{Baumann}},
  \bibinfo{author}{\bibfnamefont{C.~P.} \bibnamefont{Lutz}},
  \bibinfo{author}{\bibfnamefont{D.~M.} \bibnamefont{Eigler}},
  \bibnamefont{and} \bibinfo{author}{\bibfnamefont{A.~J.}
  \bibnamefont{Heinrich}}, \bibinfo{journal}{Science}
  \textbf{\bibinfo{volume}{335}}, \bibinfo{pages}{196} (\bibinfo{year}{2012}).

\bibitem[{\citenamefont{Miyamachi et~al.}(2013)\citenamefont{Miyamachi, Schuh,
  M\"{a}rkl, Bresch, Balashov, St\"{o}hr, Karlewski, Andr\'{e}, Marthaler,
  Hoffmann et~al.}}]{miy13}
\bibinfo{author}{\bibfnamefont{T.}~\bibnamefont{Miyamachi}},
  \bibinfo{author}{\bibfnamefont{T.}~\bibnamefont{Schuh}},
  \bibinfo{author}{\bibfnamefont{T.}~\bibnamefont{M\"{a}rkl}},
  \bibinfo{author}{\bibfnamefont{C.}~\bibnamefont{Bresch}},
  \bibinfo{author}{\bibfnamefont{T.}~\bibnamefont{Balashov}},
  \bibinfo{author}{\bibfnamefont{A.}~\bibnamefont{St\"{o}hr}},
  \bibinfo{author}{\bibfnamefont{C.}~\bibnamefont{Karlewski}},
  \bibinfo{author}{\bibfnamefont{S.}~\bibnamefont{Andr\'{e}}},
  \bibinfo{author}{\bibfnamefont{M.}~\bibnamefont{Marthaler}},
  \bibinfo{author}{\bibfnamefont{M.}~\bibnamefont{Hoffmann}},
  \bibnamefont{et~al.}, \bibinfo{journal}{Nature}
  \textbf{\bibinfo{volume}{503}}, \bibinfo{pages}{242} (\bibinfo{year}{2013}).

\bibitem[{\citenamefont{Carra et~al.}(1993)\citenamefont{Carra, Thole,
  Altarelli, and Wang}}]{car93}
\bibinfo{author}{\bibfnamefont{P.}~\bibnamefont{Carra}},
  \bibinfo{author}{\bibfnamefont{B.~T.} \bibnamefont{Thole}},
  \bibinfo{author}{\bibfnamefont{M.}~\bibnamefont{Altarelli}},
  \bibnamefont{and} \bibinfo{author}{\bibfnamefont{X.}~\bibnamefont{Wang}},
  \bibinfo{journal}{Phys. Rev. Lett.} \textbf{\bibinfo{volume}{70}},
  \bibinfo{pages}{694} (\bibinfo{year}{1993}).

\bibitem[{\citenamefont{Thole et~al.}(1992)\citenamefont{Thole, Carra, Sette,
  and van~der Laan}}]{tho92}
\bibinfo{author}{\bibfnamefont{B.~T.} \bibnamefont{Thole}},
  \bibinfo{author}{\bibfnamefont{P.}~\bibnamefont{Carra}},
  \bibinfo{author}{\bibfnamefont{F.}~\bibnamefont{Sette}}, \bibnamefont{and}
  \bibinfo{author}{\bibfnamefont{G.}~\bibnamefont{van~der Laan}},
  \bibinfo{journal}{Phys. Rev. Lett.} \textbf{\bibinfo{volume}{68}},
  \bibinfo{pages}{1943} (\bibinfo{year}{1992}).

\bibitem[{\citenamefont{Gambardella et~al.}(2002)\citenamefont{Gambardella,
  Dhesi, Gardonio, Grazioli, Ohresser, and Carbone}}]{gam02}
\bibinfo{author}{\bibfnamefont{P.}~\bibnamefont{Gambardella}},
  \bibinfo{author}{\bibfnamefont{S.~S.} \bibnamefont{Dhesi}},
  \bibinfo{author}{\bibfnamefont{S.}~\bibnamefont{Gardonio}},
  \bibinfo{author}{\bibfnamefont{C.}~\bibnamefont{Grazioli}},
  \bibinfo{author}{\bibfnamefont{P.}~\bibnamefont{Ohresser}}, \bibnamefont{and}
  \bibinfo{author}{\bibfnamefont{C.}~\bibnamefont{Carbone}},
  \bibinfo{journal}{Phys. Rev. Lett.} \textbf{\bibinfo{volume}{88}},
  \bibinfo{pages}{047202} (\bibinfo{year}{2002}).

\bibitem[{\citenamefont{Gambardella et~al.}(2003)\citenamefont{Gambardella,
  Rusponi, Veronese, Dhesi, Grazioli, Dallmeyer, Cabria, Zeller, Dederichs,
  Kern et~al.}}]{gam03}
\bibinfo{author}{\bibfnamefont{P.}~\bibnamefont{Gambardella}},
  \bibinfo{author}{\bibfnamefont{S.}~\bibnamefont{Rusponi}},
  \bibinfo{author}{\bibfnamefont{M.}~\bibnamefont{Veronese}},
  \bibinfo{author}{\bibfnamefont{S.~S.} \bibnamefont{Dhesi}},
  \bibinfo{author}{\bibfnamefont{C.}~\bibnamefont{Grazioli}},
  \bibinfo{author}{\bibfnamefont{A.}~\bibnamefont{Dallmeyer}},
  \bibinfo{author}{\bibfnamefont{I.}~\bibnamefont{Cabria}},
  \bibinfo{author}{\bibfnamefont{R.}~\bibnamefont{Zeller}},
  \bibinfo{author}{\bibfnamefont{P.~H.} \bibnamefont{Dederichs}},
  \bibinfo{author}{\bibfnamefont{K.}~\bibnamefont{Kern}}, \bibnamefont{et~al.},
  \bibinfo{journal}{Science} \textbf{\bibinfo{volume}{300}},
  \bibinfo{pages}{1130} (\bibinfo{year}{2003}).

\bibitem[{\citenamefont{Bryant et~al.}(2013)\citenamefont{Bryant, Spinelli,
  Wagenaar, Gerrits, and Otte}}]{Bryant13}
\bibinfo{author}{\bibfnamefont{B.}~\bibnamefont{Bryant}},
  \bibinfo{author}{\bibfnamefont{A.}~\bibnamefont{Spinelli}},
  \bibinfo{author}{\bibfnamefont{J.~J.~T.} \bibnamefont{Wagenaar}},
  \bibinfo{author}{\bibfnamefont{M.}~\bibnamefont{Gerrits}}, \bibnamefont{and}
  \bibinfo{author}{\bibfnamefont{A.~F.} \bibnamefont{Otte}},
  \bibinfo{journal}{Phys. Rev. Lett.} \textbf{\bibinfo{volume}{111}},
  \bibinfo{pages}{127203} (\bibinfo{year}{2013}).

\bibitem[{\citenamefont{Chen et~al.}(2008)\citenamefont{Chen, Fu, Ji, Zhang,
  Cheng, Ma, Zou, Duan, Jia, and Xue}}]{che08}
\bibinfo{author}{\bibfnamefont{X.}~\bibnamefont{Chen}},
  \bibinfo{author}{\bibfnamefont{Y.-S.} \bibnamefont{Fu}},
  \bibinfo{author}{\bibfnamefont{S.-H.} \bibnamefont{Ji}},
  \bibinfo{author}{\bibfnamefont{T.}~\bibnamefont{Zhang}},
  \bibinfo{author}{\bibfnamefont{P.}~\bibnamefont{Cheng}},
  \bibinfo{author}{\bibfnamefont{X.-C.} \bibnamefont{Ma}},
  \bibinfo{author}{\bibfnamefont{X.-L.} \bibnamefont{Zou}},
  \bibinfo{author}{\bibfnamefont{W.-H.} \bibnamefont{Duan}},
  \bibinfo{author}{\bibfnamefont{J.-F.} \bibnamefont{Jia}}, \bibnamefont{and}
  \bibinfo{author}{\bibfnamefont{Q.-K.} \bibnamefont{Xue}},
  \bibinfo{journal}{Phys. Rev. Lett.} \textbf{\bibinfo{volume}{101}},
  \bibinfo{pages}{197208} (\bibinfo{year}{2008}).

\bibitem[{\citenamefont{Choi and Gupta}(2014)}]{Choi14}
\bibinfo{author}{\bibfnamefont{T.}~\bibnamefont{Choi}} \bibnamefont{and}
  \bibinfo{author}{\bibfnamefont{J.~A.} \bibnamefont{Gupta}},
  \bibinfo{journal}{J. Phys.: Condens. Matter} \textbf{\bibinfo{volume}{26}},
  \bibinfo{pages}{394009} (\bibinfo{year}{2014}).

\bibitem[{\citenamefont{Donati et~al.}(2013)\citenamefont{Donati, Dubout,
  Aut\`{e}s, Patthey, Calleja, Gambardella, Yazyev, and Brune}}]{don13}
\bibinfo{author}{\bibfnamefont{F.}~\bibnamefont{Donati}},
  \bibinfo{author}{\bibfnamefont{Q.}~\bibnamefont{Dubout}},
  \bibinfo{author}{\bibfnamefont{G.}~\bibnamefont{Aut\`{e}s}},
  \bibinfo{author}{\bibfnamefont{F.}~\bibnamefont{Patthey}},
  \bibinfo{author}{\bibfnamefont{F.}~\bibnamefont{Calleja}},
  \bibinfo{author}{\bibfnamefont{P.}~\bibnamefont{Gambardella}},
  \bibinfo{author}{\bibfnamefont{O.~V.} \bibnamefont{Yazyev}},
  \bibnamefont{and} \bibinfo{author}{\bibfnamefont{H.}~\bibnamefont{Brune}},
  \bibinfo{journal}{Phys. Rev. Lett.} \textbf{\bibinfo{volume}{111}},
  \bibinfo{pages}{236801} (\bibinfo{year}{2013}).

\bibitem[{\citenamefont{Heinrich et~al.}(2004)\citenamefont{Heinrich, Gupta,
  Lutz, and Eigler}}]{hei04}
\bibinfo{author}{\bibfnamefont{A.~J.} \bibnamefont{Heinrich}},
  \bibinfo{author}{\bibfnamefont{J.~A.} \bibnamefont{Gupta}},
  \bibinfo{author}{\bibfnamefont{C.~P.} \bibnamefont{Lutz}}, \bibnamefont{and}
  \bibinfo{author}{\bibfnamefont{D.~M.} \bibnamefont{Eigler}},
  \bibinfo{journal}{Science} \textbf{\bibinfo{volume}{306}},
  \bibinfo{pages}{466} (\bibinfo{year}{2004}).

\bibitem[{\citenamefont{Heinrich et~al.}(2013)\citenamefont{Heinrich, Braun,
  Pascual, and Franke}}]{hei13}
\bibinfo{author}{\bibfnamefont{B.~W.} \bibnamefont{Heinrich}},
  \bibinfo{author}{\bibfnamefont{L.}~\bibnamefont{Braun}},
  \bibinfo{author}{\bibfnamefont{J.~I.} \bibnamefont{Pascual}},
  \bibnamefont{and} \bibinfo{author}{\bibfnamefont{K.~J.}
  \bibnamefont{Franke}}, \bibinfo{journal}{Nature Physics}
  \textbf{\bibinfo{volume}{9}}, \bibinfo{pages}{765} (\bibinfo{year}{2013}).

\bibitem[{\citenamefont{Hirjibehedin et~al.}(2007)\citenamefont{Hirjibehedin,
  Lin, Otte, Ternes, Lutz, Jones, and Heinrich}}]{hir07}
\bibinfo{author}{\bibfnamefont{C.~F.} \bibnamefont{Hirjibehedin}},
  \bibinfo{author}{\bibfnamefont{C.~Y.} \bibnamefont{Lin}},
  \bibinfo{author}{\bibfnamefont{A.~F.} \bibnamefont{Otte}},
  \bibinfo{author}{\bibfnamefont{M.}~\bibnamefont{Ternes}},
  \bibinfo{author}{\bibfnamefont{C.~P.} \bibnamefont{Lutz}},
  \bibinfo{author}{\bibfnamefont{B.~A.} \bibnamefont{Jones}}, \bibnamefont{and}
  \bibinfo{author}{\bibfnamefont{A.~J.} \bibnamefont{Heinrich}},
  \bibinfo{journal}{Science} \textbf{\bibinfo{volume}{317}},
  \bibinfo{pages}{1199} (\bibinfo{year}{2007}).

\bibitem[{\citenamefont{Hirjibehedin et~al.}(2006)\citenamefont{Hirjibehedin,
  Lutz, and Heinrich}}]{hir06}
\bibinfo{author}{\bibfnamefont{C.~F.} \bibnamefont{Hirjibehedin}},
  \bibinfo{author}{\bibfnamefont{C.~P.} \bibnamefont{Lutz}}, \bibnamefont{and}
  \bibinfo{author}{\bibfnamefont{A.~J.} \bibnamefont{Heinrich}},
  \bibinfo{journal}{Science} \textbf{\bibinfo{volume}{312}},
  \bibinfo{pages}{1021} (\bibinfo{year}{2006}).

\bibitem[{\citenamefont{Kahle et~al.}(2012)\citenamefont{Kahle, Deng,
  Malinowski, Tonnoir, Forment-Aliaga, Thontasen, Rinke, Le, Turkowski, Rahman
  et~al.}}]{kah12}
\bibinfo{author}{\bibfnamefont{S.}~\bibnamefont{Kahle}},
  \bibinfo{author}{\bibfnamefont{Z.}~\bibnamefont{Deng}},
  \bibinfo{author}{\bibfnamefont{N.}~\bibnamefont{Malinowski}},
  \bibinfo{author}{\bibfnamefont{C.}~\bibnamefont{Tonnoir}},
  \bibinfo{author}{\bibfnamefont{A.}~\bibnamefont{Forment-Aliaga}},
  \bibinfo{author}{\bibfnamefont{N.}~\bibnamefont{Thontasen}},
  \bibinfo{author}{\bibfnamefont{G.}~\bibnamefont{Rinke}},
  \bibinfo{author}{\bibfnamefont{D.}~\bibnamefont{Le}},
  \bibinfo{author}{\bibfnamefont{V.}~\bibnamefont{Turkowski}},
  \bibinfo{author}{\bibfnamefont{T.~S.} \bibnamefont{Rahman}},
  \bibnamefont{et~al.}, \bibinfo{journal}{Nano Letters}
  \textbf{\bibinfo{volume}{12}}, \bibinfo{pages}{518} (\bibinfo{year}{2012}).

\bibitem[{\citenamefont{Khajetoorians et~al.}(2010)\citenamefont{Khajetoorians,
  Chilian, Wiebe, Schuwalow, Lechermann, and Wiesendanger}}]{kha10}
\bibinfo{author}{\bibfnamefont{A.~A.} \bibnamefont{Khajetoorians}},
  \bibinfo{author}{\bibfnamefont{B.}~\bibnamefont{Chilian}},
  \bibinfo{author}{\bibfnamefont{J.}~\bibnamefont{Wiebe}},
  \bibinfo{author}{\bibfnamefont{S.}~\bibnamefont{Schuwalow}},
  \bibinfo{author}{\bibfnamefont{F.}~\bibnamefont{Lechermann}},
  \bibnamefont{and}
  \bibinfo{author}{\bibfnamefont{R.}~\bibnamefont{Wiesendanger}},
  \bibinfo{journal}{Nature} \textbf{\bibinfo{volume}{467}},
  \bibinfo{pages}{1084} (\bibinfo{year}{2010}).

\bibitem[{\citenamefont{Khajetoorians et~al.}(2011)\citenamefont{Khajetoorians,
  Lounis, Chilian, Costa, Zhou, Mills, Wiebe, and Wiesendanger}}]{kha11}
\bibinfo{author}{\bibfnamefont{A.~A.} \bibnamefont{Khajetoorians}},
  \bibinfo{author}{\bibfnamefont{S.}~\bibnamefont{Lounis}},
  \bibinfo{author}{\bibfnamefont{B.}~\bibnamefont{Chilian}},
  \bibinfo{author}{\bibfnamefont{A.~T.} \bibnamefont{Costa}},
  \bibinfo{author}{\bibfnamefont{L.}~\bibnamefont{Zhou}},
  \bibinfo{author}{\bibfnamefont{D.~L.} \bibnamefont{Mills}},
  \bibinfo{author}{\bibfnamefont{J.}~\bibnamefont{Wiebe}}, \bibnamefont{and}
  \bibinfo{author}{\bibfnamefont{R.}~\bibnamefont{Wiesendanger}},
  \bibinfo{journal}{Phys. Rev. Lett.} \textbf{\bibinfo{volume}{106}},
  \bibinfo{pages}{037205} (\bibinfo{year}{2011}).

\bibitem[{\citenamefont{Khajetoorians et~al.}(2013)\citenamefont{Khajetoorians,
  Schlenk, Schweflinghaus, dos Santos~Dias, Steinbrecher, Bouhassoune, Lounis,
  Wiebe, and Wiesendanger}}]{kha13}
\bibinfo{author}{\bibfnamefont{A.~A.} \bibnamefont{Khajetoorians}},
  \bibinfo{author}{\bibfnamefont{T.}~\bibnamefont{Schlenk}},
  \bibinfo{author}{\bibfnamefont{B.}~\bibnamefont{Schweflinghaus}},
  \bibinfo{author}{\bibfnamefont{M.}~\bibnamefont{dos Santos~Dias}},
  \bibinfo{author}{\bibfnamefont{M.}~\bibnamefont{Steinbrecher}},
  \bibinfo{author}{\bibfnamefont{M.}~\bibnamefont{Bouhassoune}},
  \bibinfo{author}{\bibfnamefont{S.}~\bibnamefont{Lounis}},
  \bibinfo{author}{\bibfnamefont{J.}~\bibnamefont{Wiebe}}, \bibnamefont{and}
  \bibinfo{author}{\bibfnamefont{R.}~\bibnamefont{Wiesendanger}},
  \bibinfo{journal}{Phys. Rev. Lett.} \textbf{\bibinfo{volume}{111}},
  \bibinfo{pages}{157204} (\bibinfo{year}{2013}).

\bibitem[{\citenamefont{Loth et~al.}(2010{\natexlab{a}})\citenamefont{Loth,
  Bergmann, Ternes, Otte, Lutz, and Heinrich}}]{lot10np}
\bibinfo{author}{\bibfnamefont{S.}~\bibnamefont{Loth}},
  \bibinfo{author}{\bibfnamefont{K.~v.} \bibnamefont{Bergmann}},
  \bibinfo{author}{\bibfnamefont{M.}~\bibnamefont{Ternes}},
  \bibinfo{author}{\bibfnamefont{A.~F.} \bibnamefont{Otte}},
  \bibinfo{author}{\bibfnamefont{C.~P.} \bibnamefont{Lutz}}, \bibnamefont{and}
  \bibinfo{author}{\bibfnamefont{A.~J.} \bibnamefont{Heinrich}},
  \bibinfo{journal}{Nat. Phys.} \textbf{\bibinfo{volume}{6}},
  \bibinfo{pages}{340} (\bibinfo{year}{2010}{\natexlab{a}}).

\bibitem[{\citenamefont{Loth et~al.}(2010{\natexlab{b}})\citenamefont{Loth,
  Etzkorn, Lutz, Eigler, and Heinrich}}]{lot10s}
\bibinfo{author}{\bibfnamefont{S.}~\bibnamefont{Loth}},
  \bibinfo{author}{\bibfnamefont{M.}~\bibnamefont{Etzkorn}},
  \bibinfo{author}{\bibfnamefont{C.~P.} \bibnamefont{Lutz}},
  \bibinfo{author}{\bibfnamefont{D.~M.} \bibnamefont{Eigler}},
  \bibnamefont{and} \bibinfo{author}{\bibfnamefont{A.~J.}
  \bibnamefont{Heinrich}}, \bibinfo{journal}{Science}
  \textbf{\bibinfo{volume}{329}}, \bibinfo{pages}{1628}
  (\bibinfo{year}{2010}{\natexlab{b}}).

\bibitem[{\citenamefont{Oberg et~al.}(2014)\citenamefont{Oberg, Reyes~Calvo,
  Delgado, Moro-Lagares, Serrate, Jacob, Fern\'{a}ndez-Rossier, and
  Hirjibehedin}}]{obe14}
\bibinfo{author}{\bibfnamefont{J.~C.} \bibnamefont{Oberg}},
  \bibinfo{author}{\bibfnamefont{M.}~\bibnamefont{Reyes~Calvo}},
  \bibinfo{author}{\bibfnamefont{F.}~\bibnamefont{Delgado}},
  \bibinfo{author}{\bibfnamefont{M.}~\bibnamefont{Moro-Lagares}},
  \bibinfo{author}{\bibfnamefont{D.}~\bibnamefont{Serrate}},
  \bibinfo{author}{\bibfnamefont{D.}~\bibnamefont{Jacob}},
  \bibinfo{author}{\bibfnamefont{J.}~\bibnamefont{Fern\'{a}ndez-Rossier}},
  \bibnamefont{and} \bibinfo{author}{\bibfnamefont{C.~F.}
  \bibnamefont{Hirjibehedin}}, \bibinfo{journal}{Nature Nanotechnology}
  \textbf{\bibinfo{volume}{9}}, \bibinfo{pages}{64} (\bibinfo{year}{2014}).

\bibitem[{\citenamefont{Otte et~al.}(2008)\citenamefont{Otte, Ternes, Bergmann,
  Loth, Brune, Lutz, Hirjibehedin, and Heinrich}}]{ott08}
\bibinfo{author}{\bibfnamefont{A.~F.} \bibnamefont{Otte}},
  \bibinfo{author}{\bibfnamefont{M.}~\bibnamefont{Ternes}},
  \bibinfo{author}{\bibfnamefont{K.~v.} \bibnamefont{Bergmann}},
  \bibinfo{author}{\bibfnamefont{S.}~\bibnamefont{Loth}},
  \bibinfo{author}{\bibfnamefont{H.}~\bibnamefont{Brune}},
  \bibinfo{author}{\bibfnamefont{C.~P.} \bibnamefont{Lutz}},
  \bibinfo{author}{\bibfnamefont{C.~F.} \bibnamefont{Hirjibehedin}},
  \bibnamefont{and} \bibinfo{author}{\bibfnamefont{A.~J.}
  \bibnamefont{Heinrich}}, \bibinfo{journal}{Nat. Phys.}
  \textbf{\bibinfo{volume}{4}}, \bibinfo{pages}{847} (\bibinfo{year}{2008}).

\bibitem[{\citenamefont{Schuh et~al.}(2012)\citenamefont{Schuh, Miyamachi,
  Gerstl, Geilhufe, Hoffmannm, Ostanin, Hergert, Ernst, and
  Wulfhekel}}]{Wulf12}
\bibinfo{author}{\bibfnamefont{T.}~\bibnamefont{Schuh}},
  \bibinfo{author}{\bibfnamefont{T.}~\bibnamefont{Miyamachi}},
  \bibinfo{author}{\bibfnamefont{S.}~\bibnamefont{Gerstl}},
  \bibinfo{author}{\bibfnamefont{M.}~\bibnamefont{Geilhufe}},
  \bibinfo{author}{\bibfnamefont{M.}~\bibnamefont{Hoffmannm}},
  \bibinfo{author}{\bibfnamefont{S.}~\bibnamefont{Ostanin}},
  \bibinfo{author}{\bibfnamefont{W.}~\bibnamefont{Hergert}},
  \bibinfo{author}{\bibfnamefont{A.}~\bibnamefont{Ernst}}, \bibnamefont{and}
  \bibinfo{author}{\bibfnamefont{W.}~\bibnamefont{Wulfhekel}},
  \bibinfo{journal}{Nano Lett.} \textbf{\bibinfo{volume}{12}},
  \bibinfo{pages}{4805} (\bibinfo{year}{2012}).

\bibitem[{\citenamefont{Spinelli et~al.}(2014)\citenamefont{Spinelli, Bryant,
  Delgado, Fern\'{a}ndez-Rossier, and Otte}}]{Spinelli14}
\bibinfo{author}{\bibfnamefont{A.}~\bibnamefont{Spinelli}},
  \bibinfo{author}{\bibfnamefont{B.}~\bibnamefont{Bryant}},
  \bibinfo{author}{\bibfnamefont{F.}~\bibnamefont{Delgado}},
  \bibinfo{author}{\bibfnamefont{J.}~\bibnamefont{Fern\'{a}ndez-Rossier}},
  \bibnamefont{and} \bibinfo{author}{\bibfnamefont{A.~F.} \bibnamefont{Otte}},
  \bibinfo{journal}{Nat. Mater.} \textbf{\bibinfo{volume}{13}},
  \bibinfo{pages}{782} (\bibinfo{year}{2014}).

\bibitem[{\citenamefont{Tsukahara et~al.}(2009)\citenamefont{Tsukahara, Noto,
  Ohara, Shiraki, Takagi, Takata, Miyawaki, Taguchi, Chainani, Shin
  et~al.}}]{tsu09}
\bibinfo{author}{\bibfnamefont{N.}~\bibnamefont{Tsukahara}},
  \bibinfo{author}{\bibfnamefont{K.~I.} \bibnamefont{Noto}},
  \bibinfo{author}{\bibfnamefont{M.}~\bibnamefont{Ohara}},
  \bibinfo{author}{\bibfnamefont{S.}~\bibnamefont{Shiraki}},
  \bibinfo{author}{\bibfnamefont{N.}~\bibnamefont{Takagi}},
  \bibinfo{author}{\bibfnamefont{Y.}~\bibnamefont{Takata}},
  \bibinfo{author}{\bibfnamefont{J.}~\bibnamefont{Miyawaki}},
  \bibinfo{author}{\bibfnamefont{M.}~\bibnamefont{Taguchi}},
  \bibinfo{author}{\bibfnamefont{A.}~\bibnamefont{Chainani}},
  \bibinfo{author}{\bibfnamefont{S.}~\bibnamefont{Shin}}, \bibnamefont{et~al.},
  \bibinfo{journal}{Phys. Rev. Lett.} \textbf{\bibinfo{volume}{102}},
  \bibinfo{pages}{167203} (\bibinfo{year}{2009}).

\bibitem[{\citenamefont{van Bergmann et~al.}(2015)\citenamefont{van Bergmann,
  Ternes, Loth, Lutz, and Heinrich}}]{Bergmann15}
\bibinfo{author}{\bibfnamefont{K.}~\bibnamefont{van Bergmann}},
  \bibinfo{author}{\bibfnamefont{T.}~\bibnamefont{Ternes}},
  \bibinfo{author}{\bibfnamefont{S.}~\bibnamefont{Loth}},
  \bibinfo{author}{\bibfnamefont{C.~P.} \bibnamefont{Lutz}}, \bibnamefont{and}
  \bibinfo{author}{\bibfnamefont{A.~J.} \bibnamefont{Heinrich}},
  \bibinfo{journal}{Phys. Rev. Lett.} \textbf{\bibinfo{volume}{114}},
  \bibinfo{pages}{076601} (\bibinfo{year}{2015}).

\bibitem[{\citenamefont{Yan et~al.}(2015)\citenamefont{Yan, Choi, Burgess,
  Rolf-Pissarczyk, and Loth}}]{Yan15}
\bibinfo{author}{\bibfnamefont{S.}~\bibnamefont{Yan}},
  \bibinfo{author}{\bibfnamefont{D.~J.} \bibnamefont{Choi}},
  \bibinfo{author}{\bibfnamefont{J.~A.~J.} \bibnamefont{Burgess}},
  \bibinfo{author}{\bibfnamefont{S.}~\bibnamefont{Rolf-Pissarczyk}},
  \bibnamefont{and} \bibinfo{author}{\bibfnamefont{S.}~\bibnamefont{Loth}},
  \bibinfo{journal}{Nano Lett.} \textbf{\bibinfo{volume}{15}},
  \bibinfo{pages}{1938} (\bibinfo{year}{2015}).

\bibitem[{\citenamefont{Abragam and Bleaney}(1970)}]{abr70}
\bibinfo{author}{\bibfnamefont{A.}~\bibnamefont{Abragam}} \bibnamefont{and}
  \bibinfo{author}{\bibfnamefont{B.}~\bibnamefont{Bleaney}},
  \emph{\bibinfo{title}{Electron paramagnetic resonance of transition ions}}
  (\bibinfo{publisher}{Clarendon Press}, \bibinfo{address}{Oxford},
  \bibinfo{year}{1970}).

\bibitem[{\citenamefont{Rudowicz and Karbowiak}(2015)}]{Rudowicz15}
\bibinfo{author}{\bibfnamefont{C.}~\bibnamefont{Rudowicz}} \bibnamefont{and}
  \bibinfo{author}{\bibfnamefont{M.}~\bibnamefont{Karbowiak}},
  \bibinfo{journal}{Coordin. Chem. Rev.} \textbf{\bibinfo{volume}{287}},
  \bibinfo{pages}{28} (\bibinfo{year}{2015}).

\bibitem[{\citenamefont{Wernsdorfer}(2010)}]{wer10}
\bibinfo{author}{\bibfnamefont{W.}~\bibnamefont{Wernsdorfer}},
  \bibinfo{journal}{Int. J. Nanotechnol.} \textbf{\bibinfo{volume}{7}},
  \bibinfo{pages}{497} (\bibinfo{year}{2010}).

\bibitem[{\citenamefont{Ternes}(2015)}]{Ternes15}
\bibinfo{author}{\bibfnamefont{M.}~\bibnamefont{Ternes}}, \bibinfo{journal}{New
  J. Phys.} \textbf{\bibinfo{volume}{17}}, \bibinfo{pages}{063016}
  (\bibinfo{year}{2015}).

\bibitem[{\citenamefont{Ellmer et~al.}(2001)\citenamefont{Ellmer, Repain,
  Rousset, Croset, Sotto, and Zeppenfeld}}]{ell01}
\bibinfo{author}{\bibfnamefont{H.}~\bibnamefont{Ellmer}},
  \bibinfo{author}{\bibfnamefont{V.}~\bibnamefont{Repain}},
  \bibinfo{author}{\bibfnamefont{S.}~\bibnamefont{Rousset}},
  \bibinfo{author}{\bibfnamefont{B.}~\bibnamefont{Croset}},
  \bibinfo{author}{\bibfnamefont{M.}~\bibnamefont{Sotto}}, \bibnamefont{and}
  \bibinfo{author}{\bibfnamefont{P.}~\bibnamefont{Zeppenfeld}},
  \bibinfo{journal}{Surf. Sci.} \textbf{\bibinfo{volume}{476}},
  \bibinfo{pages}{95} (\bibinfo{year}{2001}).

\bibitem[{\citenamefont{Brune}(1998)}]{bru98}
\bibinfo{author}{\bibfnamefont{H.}~\bibnamefont{Brune}},
  \bibinfo{journal}{Surf. Sci. Rep.} \textbf{\bibinfo{volume}{31}},
  \bibinfo{pages}{125} (\bibinfo{year}{1998}).

\bibitem[{\citenamefont{Shick et~al.}(2009)\citenamefont{Shick, M{\'a}ca, and
  Lichtenstein}}]{shi09}
\bibinfo{author}{\bibfnamefont{A.~B.} \bibnamefont{Shick}},
  \bibinfo{author}{\bibfnamefont{F.}~\bibnamefont{M{\'a}ca}}, \bibnamefont{and}
  \bibinfo{author}{\bibfnamefont{A.~I.} \bibnamefont{Lichtenstein}},
  \bibinfo{journal}{Phys. Rev. B} \textbf{\bibinfo{volume}{79}},
  \bibinfo{pages}{172409} (\bibinfo{year}{2009}).

\bibitem[{\citenamefont{Kresse and Furthm{\"u}ller}(1996)}]{kre96}
\bibinfo{author}{\bibfnamefont{G.}~\bibnamefont{Kresse}} \bibnamefont{and}
  \bibinfo{author}{\bibfnamefont{J.}~\bibnamefont{Furthm{\"u}ller}},
  \bibinfo{journal}{Comput. Mater. Sci.} \textbf{\bibinfo{volume}{6}},
  \bibinfo{pages}{15} (\bibinfo{year}{1996}).

\bibitem[{\citenamefont{Kresse and Hafner}(1993)}]{kre93}
\bibinfo{author}{\bibfnamefont{G.}~\bibnamefont{Kresse}} \bibnamefont{and}
  \bibinfo{author}{\bibfnamefont{J.}~\bibnamefont{Hafner}},
  \bibinfo{journal}{Phys. Rev. B} \textbf{\bibinfo{volume}{47}},
  \bibinfo{pages}{558} (\bibinfo{year}{1993}).

\bibitem[{\citenamefont{Kresse and Joubert}(1999)}]{kre99}
\bibinfo{author}{\bibfnamefont{G.}~\bibnamefont{Kresse}} \bibnamefont{and}
  \bibinfo{author}{\bibfnamefont{D.}~\bibnamefont{Joubert}},
  \bibinfo{journal}{Phys. Rev. B} \textbf{\bibinfo{volume}{59}},
  \bibinfo{pages}{1758} (\bibinfo{year}{1999}).

\bibitem[{\citenamefont{Wimmer et~al.}(1981)\citenamefont{Wimmer, Krakauer,
  Weinert, and Freeman}}]{wim81}
\bibinfo{author}{\bibfnamefont{E.}~\bibnamefont{Wimmer}},
  \bibinfo{author}{\bibfnamefont{H.}~\bibnamefont{Krakauer}},
  \bibinfo{author}{\bibfnamefont{M.}~\bibnamefont{Weinert}}, \bibnamefont{and}
  \bibinfo{author}{\bibfnamefont{A.~J.} \bibnamefont{Freeman}},
  \bibinfo{journal}{Phys. Rev. B} \textbf{\bibinfo{volume}{24}},
  \bibinfo{pages}{864} (\bibinfo{year}{1981}).

\bibitem[{\citenamefont{Shick et~al.}(1997)\citenamefont{Shick, Novikov, and
  Freeman}}]{shi97}
\bibinfo{author}{\bibfnamefont{A.~B.} \bibnamefont{Shick}},
  \bibinfo{author}{\bibfnamefont{D.~L.} \bibnamefont{Novikov}},
  \bibnamefont{and} \bibinfo{author}{\bibfnamefont{A.~J.}
  \bibnamefont{Freeman}}, \bibinfo{journal}{Phys. Rev. B}
  \textbf{\bibinfo{volume}{56}}, \bibinfo{pages}{R14259}
  (\bibinfo{year}{1997}).

\bibitem[{\citenamefont{Shick and Pickett}(2001)}]{shi01}
\bibinfo{author}{\bibfnamefont{A.~B.} \bibnamefont{Shick}} \bibnamefont{and}
  \bibinfo{author}{\bibfnamefont{W.~E.} \bibnamefont{Pickett}},
  \bibinfo{journal}{Phys. Rev. Lett.} \textbf{\bibinfo{volume}{86}},
  \bibinfo{pages}{300} (\bibinfo{year}{2001}).

\bibitem[{\citenamefont{Bruno}(1989)}]{bru89}
\bibinfo{author}{\bibfnamefont{P.}~\bibnamefont{Bruno}},
  \bibinfo{journal}{Phys. Rev. B} \textbf{\bibinfo{volume}{39}},
  \bibinfo{pages}{865} (\bibinfo{year}{1989}).

\bibitem[{\citenamefont{Ederer et~al.}(2003{\natexlab{a}})\citenamefont{Ederer,
  Komelj, Davenport, and F\"{a}hnle}}]{ede03}
\bibinfo{author}{\bibfnamefont{C.}~\bibnamefont{Ederer}},
  \bibinfo{author}{\bibfnamefont{M.}~\bibnamefont{Komelj}},
  \bibinfo{author}{\bibfnamefont{J.~W.} \bibnamefont{Davenport}},
  \bibnamefont{and}
  \bibinfo{author}{\bibfnamefont{M.}~\bibnamefont{F\"{a}hnle}},
  \bibinfo{journal}{J. Electron Spectrosc.} \textbf{\bibinfo{volume}{130}},
  \bibinfo{pages}{97} (\bibinfo{year}{2003}{\natexlab{a}}).

\bibitem[{\citenamefont{Ederer et~al.}(2003{\natexlab{b}})\citenamefont{Ederer,
  Komelj, and F\"{a}hnle}}]{ede03b}
\bibinfo{author}{\bibfnamefont{C.}~\bibnamefont{Ederer}},
  \bibinfo{author}{\bibfnamefont{M.}~\bibnamefont{Komelj}}, \bibnamefont{and}
  \bibinfo{author}{\bibfnamefont{M.}~\bibnamefont{F\"{a}hnle}},
  \bibinfo{journal}{Phys. Rev. B} \textbf{\bibinfo{volume}{68}},
  \bibinfo{pages}{052402} (\bibinfo{year}{2003}{\natexlab{b}}).

\bibitem[{\citenamefont{Pryce}(1950)}]{pry50}
\bibinfo{author}{\bibfnamefont{M.~H.~L.} \bibnamefont{Pryce}},
  \bibinfo{journal}{Proc. Phys. Soc. A} \textbf{\bibinfo{volume}{63}},
  \bibinfo{pages}{25} (\bibinfo{year}{1950}).

\bibitem[{\citenamefont{Schweflinghaus
  et~al.}(2014)\citenamefont{Schweflinghaus, dos Santos~Dias, Costa, and
  Lounis}}]{lounis14}
\bibinfo{author}{\bibfnamefont{B.}~\bibnamefont{Schweflinghaus}},
  \bibinfo{author}{\bibfnamefont{M.}~\bibnamefont{dos Santos~Dias}},
  \bibinfo{author}{\bibfnamefont{A.~T.} \bibnamefont{Costa}}, \bibnamefont{and}
  \bibinfo{author}{\bibfnamefont{S.}~\bibnamefont{Lounis}},
  \bibinfo{journal}{Phys. Rev. B} \textbf{\bibinfo{volume}{89}},
  \bibinfo{pages}{235439} (\bibinfo{year}{2014}).

\bibitem[{\citenamefont{\c{S}a\c{s}io\u{g}lu
  et~al.}(2011)\citenamefont{\c{S}a\c{s}io\u{g}lu, Friedrich, and
  Bl\"{u}gel}}]{Sasioglu15}
\bibinfo{author}{\bibfnamefont{E.}~\bibnamefont{\c{S}a\c{s}io\u{g}lu}},
  \bibinfo{author}{\bibfnamefont{C.}~\bibnamefont{Friedrich}},
  \bibnamefont{and}
  \bibinfo{author}{\bibfnamefont{S.}~\bibnamefont{Bl\"{u}gel}},
  \bibinfo{journal}{Phys. Rev. B} \textbf{\bibinfo{volume}{83}},
  \bibinfo{pages}{121101(R)} (\bibinfo{year}{2011}).

\bibitem[{\citenamefont{Corradini et~al.}(2012)\citenamefont{Corradini, Ghirri,
  Garlatti, Biagi, De~Renzi, del Pennino, Bellini, Carretta, Santini, Timco
  et~al.}}]{cor12}
\bibinfo{author}{\bibfnamefont{V.}~\bibnamefont{Corradini}},
  \bibinfo{author}{\bibfnamefont{A.}~\bibnamefont{Ghirri}},
  \bibinfo{author}{\bibfnamefont{E.}~\bibnamefont{Garlatti}},
  \bibinfo{author}{\bibfnamefont{R.}~\bibnamefont{Biagi}},
  \bibinfo{author}{\bibfnamefont{V.}~\bibnamefont{De~Renzi}},
  \bibinfo{author}{\bibfnamefont{U.}~\bibnamefont{del Pennino}},
  \bibinfo{author}{\bibfnamefont{V.}~\bibnamefont{Bellini}},
  \bibinfo{author}{\bibfnamefont{S.}~\bibnamefont{Carretta}},
  \bibinfo{author}{\bibfnamefont{P.}~\bibnamefont{Santini}},
  \bibinfo{author}{\bibfnamefont{G.}~\bibnamefont{Timco}},
  \bibnamefont{et~al.}, \bibinfo{journal}{Adv. Funct. Mater.}
  \textbf{\bibinfo{volume}{22}}, \bibinfo{pages}{3706} (\bibinfo{year}{2012}).

\bibitem[{\citenamefont{Stepanow et~al.}(2010)\citenamefont{Stepanow, Mugarza,
  Ceballos, Moras, Cezar, Carbone, and Gambardella}}]{ste10}
\bibinfo{author}{\bibfnamefont{S.}~\bibnamefont{Stepanow}},
  \bibinfo{author}{\bibfnamefont{A.}~\bibnamefont{Mugarza}},
  \bibinfo{author}{\bibfnamefont{G.}~\bibnamefont{Ceballos}},
  \bibinfo{author}{\bibfnamefont{P.}~\bibnamefont{Moras}},
  \bibinfo{author}{\bibfnamefont{J.~C.} \bibnamefont{Cezar}},
  \bibinfo{author}{\bibfnamefont{C.}~\bibnamefont{Carbone}}, \bibnamefont{and}
  \bibinfo{author}{\bibfnamefont{P.}~\bibnamefont{Gambardella}},
  \bibinfo{journal}{Phys. Rev. B} \textbf{\bibinfo{volume}{82}},
  \bibinfo{pages}{014405} (\bibinfo{year}{2010}).

\end{thebibliography}

\end{document}